\documentclass[aps,pra,longbibliography,amsmath,amssymb,reprint,superscriptaddress,]{revtex4-1}

\usepackage{graphicx}
\usepackage{bm}
\usepackage{times}
\usepackage{amsmath}
\usepackage{bbold}
\usepackage[english]{babel}
\usepackage{color}
\usepackage[colorlinks]{hyperref}
\usepackage[normalem]{ulem}
\hypersetup{
colorlinks=true,
linkcolor=blue,          
citecolor=blue,          
filecolor=magenta,       
urlcolor=cyan
}

\begin{document}

\title{Polarization dependent scattering in cavity optomagnonics}
\author{J. A. Haigh}

\email{jh877@cam.ac.uk}
\affiliation{Hitachi Cambridge Laboratory, Cambridge, CB3 0HE, United Kingdom}
\author{A. Nunnenkamp}
\affiliation{School of Physics and Astronomy and Centre for the Mathematics and Theoretical Physics of Quantum Non-Equilibrium Systems, University of Nottingham, Nottingham, NG7 2RD, United Kingdom}
\affiliation{Cavendish Laboratory, University of Cambridge, Cambridge CB3 0HE, United Kingdom}

\author{A. J. Ramsay}
\affiliation{Hitachi Cambridge Laboratory, Cambridge, CB3 0HE, United Kingdom}

\date{\today}

\begin{abstract}
The polarization dependence of magnon-photon scattering in an optical microcavity is reported. Due to the short cavity length,  the mode-matching conditions found in previously explored, large path-length whispering gallery resonators are absent. Nonetheless, for cross-polarized scattering a strong and broadband suppression of one side-band is observed. This arises due to an interference between the Faraday and second-order Cotton-Mouton effects. To fully account for the suppression of the cross-polarized scattering, it is necessary to consider the squeezing of magnon modes intrinsic to thin-film geometry. A co-polarized scattering due to Cotton-Mouton effect is also observed. In addition, the magnon modes involved are identified as Damon-Eshbach surface modes, whose non-reciprocal propagation could be exploited in devices applications. This work experimentally demonstrates the important role of second order Cotton-Mouton effect for optomagnonic devices. 
\end{abstract}

\maketitle

The enhancement of the interaction between ferromagnetic magnons and optical photons has been recently explored \cite{osada_cavity_2016,zhang_optomagnonic_2016,haigh_triple-resonant_2016} towards efficient microwave-optical conversion \cite{hisatomi_bidirectional_2016,lambert_coherent_2020}, and low power optical driving of magnonic devices \cite{zhu_inverse_2021}. In combination with other experiments, these studies have demonstrated that the use of electromagnetic cavities to significantly increase the coupling of magnons to both microwave \cite{huebl_high_2013,zhang_strongly_2014,tabuchi_hybridizing_2014} and optical \cite{zhu_waveguide_2020-2,haigh_subpicoliter_2020} photons \cite{rameshti_cavity_2021}.

One of the original motivations for this work has been to transfer ideas from the field of optomechanics \cite{aspelmeyer_cavity_2014} to magnetic systems, replacing the collective mechanical mode of a solid object with the magnetic resonance mode of a ferromagnetic material. For linearized magnon modes, many ideas carry over directly \cite{viola_kusminskiy_coupled_2016}, such as dynamical cooling \cite{sharma_optical_2018,bittencourt_magnon_2019}, and coherent optical driving \cite{simic_coherent_2020}. However, there are key differences between the two systems, the most important being the role that the optical polarization plays in the scattering \cite{haigh_triple-resonant_2016}. This is highlighted by the single side-band magnon scattering observed in whispering gallery mode devices \cite{osada_cavity_2016,zhang_optomagnonic_2016,haigh_triple-resonant_2016}, with only Stokes or anti-Stokes scattering dependent on the polarization of the optical pump. 
However, to alleviate the poor mode overlap and large cavity mode volume that result in low magnon-photon coupling rates, there has been a shift to other cavity geometries, e.g. waveguide \cite{zhu_waveguide_2020-2} and microcavity devices \cite{haigh_subpicoliter_2020}.

\begin{figure}%
\includegraphics[width=\columnwidth]{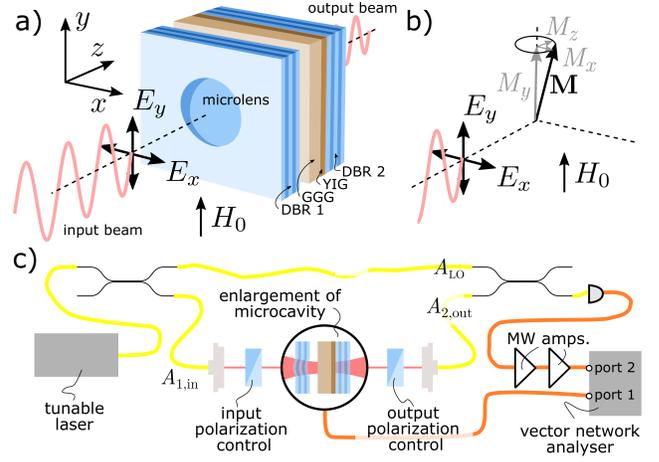}%
\caption{(a) Schematic of optomagnonic cavity. 
The YIG layer is mounted on a flat dielectric mirror (DBR 2). The cavity is completed by second mirror with input micro-lens on the inner surface (DBR 1).  (b) The direction of the applied magnetic field $H_0$ is along $\hat{y}$, with magnetization precession resulting in small components $M_z$ and $M_x$ 
(c) Experimental setup for homodyne detection measurements.}%
\label{schematic}%
\end{figure}

In this letter, we report measurements of the polarization dependent magnon light scattering in a magneto-optical microcavity. The cross-polarized scattering is close to the triple resonance condition, and shows strong side-band asymmetry similar to the WGM resonators \cite{haigh_triple-resonant_2016}, despite the difference in cavity geometry. In the case of WGM resonator, the strong side-band suppression can be explained solely in terms of the first-order Faraday interaction in combination with a mode matching condition \cite{sharma_light_2017}. Here, the role of mode-matching is weak, and the asymmetry of the cross-polarized scattering  arises from an interference in the magneto-optical coupling between the first order Faraday term, and an effective first-order term that comes from mixing of the dc and ac magnetization by the Cotton-Mouton effect \cite{liu_optomagnonics_2016}. In addition, a co-polarized magnon-scattering is also observed. This is due to the Cotton-Mouton effect alone and involves a single optical resonance, with similarities to the standard optomechanical system.
The simple Fabry-Perot-like device geometry enables an experimental study of the different polarization dependent magnon scattering based solely on the symmetry of the magneto-optical tensors. This contrasts with more complex photonic structures, where a theoretical understanding of this interference is difficult \cite{graf_design_2020}, and often the analysis is simplified by neglecting the Cotton-Mouton effect \cite{viola_kusminskiy_coupled_2016}.
Furthermore, we highlight one way in which non-reciprocal device operation can be directly embedded in optomagnonic devices, by showing that it is the chiral Damon-Eshbach magnon modes which give the largest response in our experiment.

A schematic of the device is shown in Fig.~\ref{schematic}(a). Details of the fabrication can be found in Ref.~\cite{haigh_subpicoliter_2020}.  It consists of an open micro-cavity \cite{barbour_tunable_2011} with an embedded layer of single crystal YIG with growth axis along $[111]$. The YIG is transferred from a GGG substrate and polymer bonded to a planar dielectric mirror surface. To provide lateral confinement of the optical field, the complementary mirror has an ion-milled micro-lens of radius 90~$\mathrm{\mu m}$ \cite{dolan_femtoliter_2010}. The mirror structure is designed for maximum reflectance at 1300\,nm.

The micro-cavity is probed in transmission with a tunable external cavity diode laser, see Fig.~\ref{schematic}(a). The orientation of device is shown, with optical axis along $\hat{z}$. The applied magnetic field and dc-magnetization are along the $\hat{y}$ direction. 
 The magnetization dynamics are driven by a microwave tone from the vector network analyzer (VNA), via an on-chip microwave strip-line antenna aligned along $y$-axis \cite{haigh_subpicoliter_2020}.

\begin{figure}%
\includegraphics[width=\columnwidth]{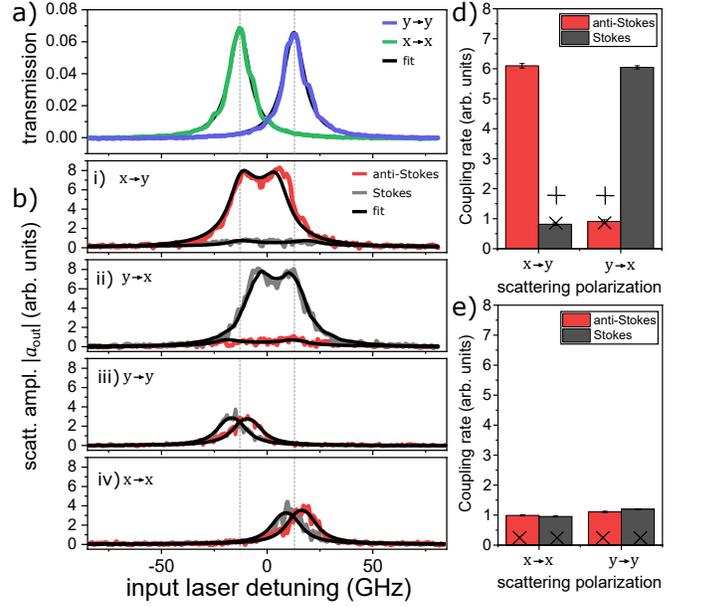}%
\caption{(a) Transmission measurement of optical cavity modes for linearly polarized light along $x$ and $y$, with fits shown in black. (b) Amplitude of magnon scattered optical signal for different input and output polarizations (i-iv). The black lines are linear regression of the lineshape given by Eq.~\ref{eq:spec} using parameters extracted from (a). (d-e) Comparison of relative amplitudes of coupling rates $\vert\mathcal{G}_{ij}^{\pm}\vert$ extracted from fits in (b). Colored bars are measurements, with modeled values as $\times$ and $+$ with and without squeezing, respectively. In these plots, the magnetic field is set to $\approx200$~mT, with microwave drive at 7.323~GHz. This is at the maximum in the magnon scattered optical signal, see Fig.~\ref{DE_modes}(a).}%
\label{laserdetuning}%
\end{figure}

The optical cavity modes have a free spectral range of 6.7~THz and a finesse of 600. A transmission measurements for linearly polarized light along $x$ and $y$ at around 1317~nm is shown in Fig.~\ref{laserdetuning}(a). The modes are linearly polarized, and a Lorentzian fit yields a splitting of 25.7\,GHz and linewidths of $\kappa_{x}= 11.9$~GHz and $\kappa_{y}=12.1$~GHz.

Next we measure the magnon-scattered optical signal via homodyne detection, using the experimental setup shown in Fig.~\ref{schematic}(c). The signal consists of Stokes and anti-Stokes components at optical frequency $\omega_\text{opt}\pm\omega_\text{m}$ where $\omega_\text{m}$ is the magnon mode frequency. A local oscillator taken from the input laser is added to the transmitted optical field using a beamsplitter, and mixed on a photodiode. The Stokes and anti-Stokes terms both result in microwave signals at frequency $\omega_\text{m}$, which interfere. As the laser frequency is swept, the local oscillator phase changes due to the imbalance between the optical path lengths, modulating the interference \cite{neuhaus_breaking_1998}.  The signal oscillates between a minimum and maximum value, from which the amplitude of the Stokes and anti-Stokes components can be extracted $|a_\textsc{s,as}| \propto \max{|V_\text{meas}|} \pm \min{|V_\text{meas}|}$. Note that the assignment of the $\pm$ to Stokes or anti-Stokes is not possible from the measurement due to the unknown signs of the signals. However,  the assignment can be made based on the laser detuning dependence of the extracted values.

The amplitudes of the measured Stokes and anti-Stokes signals are shown in Fig.~\ref{laserdetuning}(b). For cross polarized scattering (i-ii), we see that depending on the input polarization, one of the side-bands is suppressed. For co-polarized scattering (iii-iv), we also observe magnon scattering around the frequency of the correspondingly polarized optical mode. In this case, the measurements are similar to those expected for a standard optomechanical system, with one optical mode. Because we are in the unresolved side-band regime ($\omega_\text{m}/(2\pi)=7.323~\mathrm{GHz}< \kappa_{x,y}$), the side-band suppression is weak. Remarkably, for cross-polarized scattering the side-band suppression is strong over the entire detuning range. This is despite a mode-splitting similar to the cavity linewidth, and the relaxation of the mode-matching conditions due to the small optical mode volume. Therefore, this suppression cannot be explained by the Faraday effect alone.
As shown theoretically in Ref.~\cite{liu_optomagnonics_2016}, side-band suppression in a cavity optomagnonic system can result from the interference of the Faraday interaction and the second order Cotton-Mouton effect. This effect has also been observed in Brillouin light scattering experiments in bulk YIG \cite{wettling_relation_1975}. In the following, we show how this arises. Additionally, we find it necessary to include the squeezing of the magnon mode.

The interaction Hamiltonian for the magneto-optical coupling of optical modes $\hat{a}_{x,y}$ and magnon mode $\hat{b}$ has the form
\begin{eqnarray}
H_\text{int}= \hbar\sum_{ij} ( \mathcal{G}_{ij}^{+} a^{\dagger}_i a_j^{} b^{} +  \mathcal{G}_{ij}^{-} a^{\dagger}_i a_j^{} b^{\dagger}),
\label{Hint_sum}
\end{eqnarray}
where
\begin{equation}
\mathcal{G}_{ij}^{\pm} =\mathcal{G}\mathcal{A}_{ij}^{\pm} \eta_{ij}^\pm.
\label{rate_int}
\end{equation}
The overlap between the three relevant modes is expressed as $\eta_{ij}^\pm =\int{  u_\text{m}^{(*)}(\mathbf{r})  {u}_{i}^{*} (\mathbf{r}) {u}_{j}^{}(\mathbf{r})dV}$ \footnote{Superscript $(*)$ indicates conjugate only for $\eta^-$}, with $u_\text{m}(\mathbf{r})$ the magnon mode function, and $\mathcal{G}$ is a material-dependent parameter. For our microcavity, we can set $\eta_{ij}^\pm \approx \eta$ as the optical mode functions $u_x(\mathbf{r})\approx u_y(\mathbf{r})$ \cite{Note2}. We can therefore understand the polarization dependent scattering in terms of the elements of $\mathcal{A}$, which is defined such that the ac-term of the dielectric constant $\varepsilon_{ij}^{ac}=\frac{1}{2}\sum_{\pm}\mathcal{A}_{ij}^{\pm}M^{\pm}$, where $M^{\pm}=M_z \pm i M_x$. The separation of the spatial mode overlap $\eta$ and dielectric-tensor-dependent matrix $\mathcal{A}$ is only possible because of the simple mode structure of the optical cavity, in contrast to, e.g. whispering gallery mode resonators \cite{haigh_triple-resonant_2016}.

The dielectric tensor can be expanded in powers of magnetization, $\varepsilon_{ij} = \varepsilon^0_{ij} +  K_{ijk} M_k + G_{ijkl} M_k M_l + ...$. Here, the linear in magnetization part $K_{ijk} = i \epsilon_{ijk} f $ is the Faraday term with Levi-Civita symbol $\epsilon_{ijk}$. The second order part with four-rank tensor $G_{ijkl}$ gives rise to the Cotton-Mouton effect (linear birefringence). Due to the strong dc magnetization $M_0$, the Cotton-Mouton tensor results in an ac-term that contributes to the one-magnon scattering $\sim M_0 M_{x,y}$, a dc-term contributing to the static birefringence proportional to $M_0^2$, and a two-magnon scattering term $\sim M^2$ \cite{hisatomi_helicity-changing_2019}. We disregard the latter two parts, as we are interested in only one magnon scattering events, and the birefringence is small compared to the cavity mode splitting. For a cubic material such as YIG, the Cotton-Mouton tensor has only three independent parameters, $G_{11}$, $G_{12}$, and $G_{44}$ \cite{nye_physical_1985}, and we define $g \equiv G_{11}-G_{12} - 2 G_{44}$ to simplify the expressions \cite{prokhorov_optical_1984}. The Gaussian modes of our optical microcavity can be treated in paraxial approximation, where only the $\hat{x},\hat{y}$ polarization components of electric-field are considered. This eliminates the $z$-components of the dielectric tensor from our analysis. 

Importantly, the effective first-order dielectric tensor contains terms in both $M_z$ and $M_x$, with different relative phases, leading to interference between the two terms \footnote{See Supplementary information}, as can be seen in the structure of
\begin{align}
\mathcal{A}^{\pm}_{ij} &= \nonumber\\
 &
\begin{bmatrix}
-\frac{\sqrt{2}g}{3}M_0  & - i (f \mp (2 G_{44} + \frac{g}{3})M_0)  \\
i(f \pm (2 G_{44} + \frac{g}{3})M_0)  & \frac{\sqrt{2}g}{3} M_0  \\
\end{bmatrix}.
\label{Aij}
\end{align}
Here, we have used the Cotton-Mouton tensor for YIG with $(x,y,z)=([10\bar{1}],[\bar{1}2\bar{1}],[111])$.

Firstly, we see that the diagonal elements of $\mathcal{G}^{\pm}_{ij}$ (proportional to those of $\mathcal{A}^{\pm}_{ij}$) are non-zero, resulting in the co-polarized scattering we observe. Secondly, due to an interference between Faraday and Cotton-Mouton terms the off-diagonal elements of $\mathcal{G}_{ij}$ are not equal, resulting
in the side-band asymmetry. Although this is the case, the interaction Hamiltonian is still Hermitian, given that $\mathcal{G}_{ij}^+=\mathcal{G}_{ji}^-$.
In the case $f = \pm(2\mathcal{G}_{44} + \frac{g}{3})M_0$, there is complete side band suppression. Due to the crystal orientation, the Cotton-Mouton effect also results in co-polarized diagonal terms, with $\mathcal{G}^-_{ii}=\mathcal{G}^+_{ii}$.

The output of the cavity at the Stokes (S) and anti-Stokes (AS) frequencies for the different polarization configuration can be calculated from the linearized dynamics using the interaction Hamiltonian and solving the quantum Langevin equations \cite{rameshti_cavity_2021}. This gives \cite{Note2,haigh_triple-resonant_2016}
\begin{equation}
|\langle \hat{a}_{i,\text{out}}^{\textsc{s/as}} \rangle|^2
= \frac{4 \mathcal{G}_{ij}^{\pm2} |\bar{a}_{j,\text{in}}|^2 |\bar{b}_\text{in}|^2 \kappa_i \kappa_j/\kappa_\textsc{fmr}}{\left(\frac{\kappa_j^2}{4}+\Delta_j^2\right) \left(\frac{\kappa_i^2}{4}+(\Delta_i\pm\omega_\text{m})^2\right)}, \label{eq:spec}
\end{equation}
where $|\bar{b}_\text{in}|$ is the input microwave amplitude, and $\Delta_i = \omega_\text{in}-\omega_i$ is the detuning of the optical input from the cavity mode frequency $\omega_i$, with amplitude $\vert\bar{a}_{i,\text{in}}\vert$, $\kappa_i$ and $\kappa_\textsc{fmr}$ are the dissipation rates of the optical and magnetic modes, respectively and $\omega_\text{m}$ is the magnon mode frequency. All parameters defining the lineshape are known from the transmission measurements, so linear regression is used to extract the overall amplitude from the data, as shown in Fig.~\ref{laserdetuning}(b). Because the efficiency of the microwave coupling is unknown, the ratio of the coupling rates, e.g. $|\mathcal{G}^-_{xy}|/|\mathcal{G}^+_{xy}|=0.13\pm0.01$, $|\mathcal{G}^-_{yx}|/|\mathcal{G}^+_{yx}|=0.15\pm0.01$ are found. We compare these values to those expected given the elements of the Cotton-Mouton tensor, $G_{44} M_0^2= -1.14\times10^{-4}$, and $g M_0^2 = 5.73\times10^{-5} $ and the Faraday coefficient $f M_0 = 3.81 \times10^{-4} $ \cite{stancil_spin_2009}, where the saturation magnetization $M_0=139$~kAm$^{-1}$. This analysis, which assumes circular precession of the magnetization, underestimates the cross-polarized side-band suppression, (marked as $+$) in Fig. \ref{laserdetuning}(d).  Note that, $\mathcal{A}$ neglects the unknown in-plane angle $\phi$ between $x$ and the high symmetry $[10\bar{1}]$ crystallographic axis. This adds an angle dependent component to the dielectric tensor with a phase that cannot interfere with the off-diagonal terms \cite{Note2}, and therefore cannot increase the sideband suppression from the values above.

To account for the enhanced side-band suppression, we consider the ellipticity of the magnetization precession in a confined magnetic structure. The ellipticity is analogous to squeezing of the amplitudes in quantum optics \cite{sharma_spin_2021}, and is introduced via a squeezing parameter $e^{2\varepsilon} = |M_x|/|M_z|$ \cite{kamra_super-poissonian_2016}. The magnon operators can be diagonalized in a different basis using the Bogoliubov transformation $\hat{b} \rightarrow \cosh(\varepsilon) \hat{b}^{} + \sinh(\varepsilon) \hat{b}^\dagger$ \cite{kamra_super-poissonian_2016,sharma_spin_2021}. This modifies the coupling rates $ \mathcal{G}_{ij}^{\pm'} = \mathcal{G}_{ij}^{\pm}\cosh(\varepsilon) -  \mathcal{G}_{ij}^{\mp}\sinh(\varepsilon)$. Based on the LLG equation in the macrospin approximation, the Kittel mode has ellipicity $|M_x|/|M_z| = \sqrt{1+M/H} \approx 1.37$ (using $\mu_0 M=0.175$~T, $\mu_0 H=0.2$~T)  \cite{sharma_spin_2021}. Likewise, the squeezing of the Damon-Eshbach mode on the low energy side of the band is similar to the Kittel mode due to the large in-plane wavevector, resulting in a demagnetizing field dominated by the out-of-plane component \cite{Note2}.
With this correction, we find $\mathcal{G}^{+'}_{xy}/\mathcal{G}^{-'}_{xy}\approx1/6$, in agreement with experiment, as shown in Fig.~\ref{laserdetuning}(d) (diagonal crosses $\times$).

The measured amplitude of the co-polarized signal is a factor of 4 stronger than expected from our calculations, shown as crosses ($\times$) in Fig.~\ref{laserdetuning}(e). We believe this is due to  magnetostrictive contributions to the Cotton-Mouton tensor \cite{prokhorov_optical_1984} which are likely to be enhanced in the YIG/GGG membrane by the weak clamping of the polymer bonding.


\begin{figure}%
\includegraphics[width=\columnwidth]{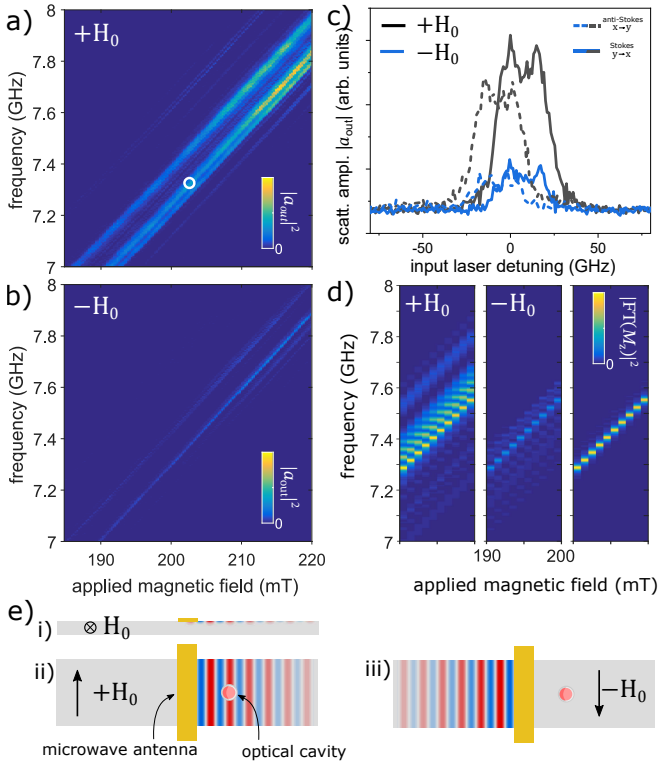}%
\caption{(a-b) Measured ferromagnetic resonance modes. The colormap shows the optical signal power of the magnon scattered signal versus applied positive (a) and negative (b) magnetic field and microwave drive frequency. The white circle in (a) shows the field/frequency of measurements in Fig.~\ref{laserdetuning}. (c) Input laser detuning dependence of magnon scattered optical amplitude for positive and negative field. (d) Micromagnetic modeling of the magnon modes at the optical cavity position (left two panels) and averaged over the entire simulation showing the Kittel mode. Colormap intensity is Fourier power of $M_z$. The absolute power depends on the magnitude of the field impulse, but the same scale is used in the left two panels to compare the positive/negative field response. (e) Schematic of the propagation of magnons away from the waveguide antenna (gold). Due to the preferential coupling to one surface mode (i), the direction is towards (ii) or away (iii) from the optical mode, dependent on the sign of the applied magnetic field.}%
\label{DE_modes}%
\end{figure}

Finally, we identify the magnon modes contributing to the optical scattering in our experiments. The power of the magnon scattered optical signal is shown in Fig.~\ref{DE_modes}, as a function of applied positive (a) and negative (b) magnetic field and microwave drive frequency. The signal strongly depends on the sign of the magnetic field. The lineshape as a function input optical if frequency is unaffected (Fig.~\ref{DE_modes}(c), hinting that this effect is not related to the optomagnonic coupling. Instead, we consider the schematic of the microwave antenna shown in Fig.~\ref{DE_modes}(e). The magnetic field is applied along the length of waveguide, such that magnons with wavevector perpendicular to the magnetic field are excited. These are the Damon-Eshbach modes \cite{eshbach_surface_1960}, which are chiral surface modes propagating in one direction on each surface. As the microwave field from the antenna decays rapidly with distance, it only couples to the mode on the top surface. For one sign of the magnetic field, the propagation of the driven mode is towards the optical cavity, whereas for the opposite magnetic field the propagation is away. This is non-reciprocal in the sense that, for one field direction, while the microwave-generated magnons propagate towards the optical cavity and participate in the optical scattering, the magnons generated by optical scattering processes are driven away from the microwave coupling line.

We use micromagnetic modeling to confirm that the Damon-Eshbach modes are observed in experiments
\cite{donahue_oommf_1999}. The model is based on the approximate geometry of the ferromagnetic layer, and simulates the ring-down of the magnetization dynamics following an impulsive field with spatial distribution calculated from the Biot-Savart law based on current uniformly distributed in the microwave antenna \cite{Note2}. The frequency response is calculated from the Fourier transform of the ring-down. This is repeated for the different values of applied magnetic field. The optical scattering response is calculated by averaging the magnetization over the optical mode volume (Fig.~\ref{DE_modes}(d) left two panels).The location of Kittel mode is found by averaging over the entire magnetic volume, resulting in the bright mode in Fig.~\ref{DE_modes}(d) right panel. We note that the micromagnetic modeling does not include the magneto-crystalline anisotropy, leading to a frequency offset in comparison to the measured data. However, the broad features are in agreement with the experiment, as shown in Fig.~\ref{DE_modes}(a-b). This includes the asymmetry in the response for positive/negative magnetic fields and the broad band of spin waves above the Kittel mode, with a dark band corresponding to the wavevectors not compatible with the microwave antenna. 

In conclusion, we have measured the polarization dependent magnon-scattering in magneto-optical micro-cavities. A co-polarized process involving one optical mode is observed. This process is stronger than expected from the Cotton-Mouton effect, and has a similar phenomenology to an optomechanical system in the non-resolved sideband regime (cavity linewidth of 12 GHz greater than the magnon frequency of $\sim$8 GHz), where the side-band suppression is weak. However, for cross-polarized scattering a strong broadband side-band suppression is observed. Interference between the first-order Faraday effect, and an effective first-order Cotton-Mouton magneto-optical effects gives rise to an intrinsic asymmetry between Stokes and anti-Stokes processes. This asymmetry is further enhanced by the intrinsic ground state squeezing naturally existing in anisotropic magnetic structures \cite{kamra_super-poissonian_2016,sharma_spin_2021}.

The cross-polarized scattering is reminiscent of the recently explored coherent scattering approach to levitated optomechanical systems \cite{delic_cavity_2019}. There, efficient mechanical cooling was achieved by alleviating the laser phase noise heating. The cross-polarization scattering in magneto-optical cooling \cite{sharma_optical_2018,bittencourt_magnon_2019} could also benefit from similar insensitivity to laser phase noise heating.  Our experiments also show that the microwave antenna couples to directed Damon-Eshbach surface modes, hinting  at the non-reciprocal functionality possible because of time-reversal symmetry breaking of the ferromagnetic magnetization \cite{jalas_what_2013}.

\section*{Acknowledgements}

This work was supported by the European Union’s Horizon 2020 Research and Innovation Programme under Grant Agreement No. 732894 (FET Proactive HOT). We are grateful for advice on micro-magnetic modeling from Pierre Roy.

\bibliography{bibliography}

\end{document}